\documentstyle[12pt]{article}
\def\beq{\begin{equation}}
\def\eeq{\end{equation}}
\newcommand{\bea}{\begin{eqnarray}}
\newcommand{\eea}{\end{eqnarray}}

\def\noi{\noindent}

\def\R{ {\rm R \kern -.31cm I \kern .15cm}}
\def\C{ {\rm C \kern -.15cm \vrule width.5pt \kern .12cm}}
\def\Z{ {\rm Z \kern -.27cm \angle \kern .02cm}}
\def\N{ {\rm N \kern -.26cm \vrule width.4pt \kern .10cm}}
\def\1{{\rm 1\mskip-4.5mu l} }
\begin{document}

\title{ Exact Duality and Bjorken Sum Rule\\
 in Heavy Quark Models \`a la Bakamjian-Thomas} 
\author{ A. Le Yaouanc, L. Oliver, O. P\`ene and J.-C. Raynal} \par 
\maketitle
\centerline{Laboratoire de Physique Th\'eorique et Hautes Energies\footnote{Laboratoire associ\'e au
Centre National de la Recherche Scientifique - URA D0063}}  
\centerline{Universit\'e de Paris XI, B\^atiment 211, 91405 Orsay Cedex, France}  

\begin{abstract}
The heavy mass limit of quark models based on the Bakamjian-Thomas cons\-truction reveals remarkable features. In addition to previously demonstrated properties of covariance and
Isgur-Wise scaling, exact duality, leading to the
Bjorken-Isgur-Wise sum rule, is proven, for the first time to our knowledge in relativistic quark models. Inelastic as well as
elastic contributions to the sum rule are then discussed in terms of ground state averages of a few number of
operators corresponding to the nonrelativistic dipole operator and various relativistic
corrections.
\end{abstract}

\vskip 1 cm \noi LPTHE Orsay 96-05 \par \noi January 1996
\newpage \pagestyle{plain}
\section{Introduction} 

In \cite{[leyaouanc1]} we have
proposed a quark model of current matrix-elements in the heavy quark
limit which has been shown to present two important features~: it is covariant and it presents the full scaling
properties of heavy quark symmetry (HQS), demons\-trated by Isgur and Wise for QCD \cite{[isgur1]}. The model is based on an old formulation of relativistic multiparticle
states, i.e. a representation of the full Poincar\'e group, proposed by Bakamjian and Thomas
\cite{[bakamjian]} and reexpressed by Osborn \cite{[osborn]}. We have given a
simpler and concise description of Poincar\'e generators and wave functions in
momentum space \cite{[leyaouanc1]}. Since some time this formulation, or variants of it,
have been used by a large variety of people to formulate quark models of form factors, especially on the
null-plane \cite{[models]}. One must stress that by boosting the states to arbitrarily
large momenta, one recovers the models directly formulated on the null-plane. In
particular, one gets null-plane wave functions with the correct kinematical limits,
unlike other approaches. \par 
It must be emphasized however that in this formulation, for finite quark masses, the
currents based on the free quark current operators are not covariant.
What we have found is that their {\it heavy quark mass limit are covariant}. One important consequence of this covariance property of the limit is that all the
models we have referred to will have essentially the same limit, whichever the frame in
which they have been formulated when $m_Q < \infty$. Then, the only remaining source of
variation consists in the choice of the mass operator or of the set of wave functions
at rest, and everything we are to say, being completely independent of such
choices, should hold for the $m_Q \to \infty$ limit of such models. \par

The implementation of covariance and heavy quark symmetry obtained in this approach
must be fully appreciated. Indeed they are not built in, neither enforced by hand. It
represents an important progress with respect to older popular models of heavy-to-heavy semi-leptonic
form factors which, on the one hand, make implicit or explicit
reference to particular frames to calculate form factors, and, which, on the other
hand, finally renounce to predict form factors except for a privileged $q^2$ value,
and inspire for instance from VMD to extrapolate form factors. Then either heavy quark
symmetry is not satisfied or it is enforced by hand. In the present approach, on the contrary,
everything results from a systematic calculation from the eigenstates of a mass
o\-pe\-rator, and a current operator. It must be also underlined that the approach, while
lea\-ding to such nontrivial properties, does not spoil the simplicity which makes quark
models so attractive. It preserves principles of quark models like a fixed number of
constituents, a three-dimensional description through ordinary wave-functions, a free
quark current operator ~. On the other hand, it is also a progress
with respect to our own older models \cite{[leyaouanc2]} which, while presenting heavy
quark symmetry, were only approximately covariant (i.e. the Isgur-Wise scaling
function $\xi$ was depending on the chosen frame). \par

In the present paper, we pursue the investigation of the model presented in
\cite{[leyaouanc1]} as regards general properties, not dependent on particular choices
of interaction or wave functions at rest~: the mass operator will remain totally
arbitrary, except of course for rotation and parity invariance. As another important
advantage of the new approach, we want now to show that it may help a lot in
understanding more physically, in the context of bound state physics, {\it sum
rules} already formulated in a field-theoretical context. Let us stress that exact
saturation of sum rules is a rather specific feature of this type of relativistic models.
Sum rules strongly rely on completeness relations. But what is precisely needed is
completeness relations for wave functions of states in {\it motion}. Such relations
are automatically provided by the construction of a unitary representation of the Poincar\'e
group. On the other hand, many phenomenological models use wave functions in motion which
are not shown to be orthonormal, e.g. to be eigenstates of an Hamiltonian or even, as
recalled above, use form factors not calculated through wave functions except for a
normalisation at some $q^2$. Then, they have no reason to satisfy sum rules. \par

In addition to demonstrating that Bjorken-Isgur-Wise (BIW) sum rule is exactly valid in our approach, we
shall also decompose the contributions of the various states to the ``physical side''
of the sum rule into different parts with clear physical meaning, clarifiying in
particular the nature of relativistic corrections and the origin of the various bounds
on $\rho^2$. \par \vskip 5 truemm

\section {Direct demonstration of duality and saturation of
the Bjorken-Isgur-Wise sum rule}

The sum rule under consideration writes, for mesons, in the transparent formulation of
Isgur and Wise \cite{[isgur2]}

\beq \rho^2 - \sum_{k} \left ( \left | \tau_{1/2}^k(1) \right |^2 + 2 \left |
\tau_{3/2}^k(1) \right |^2 \right ) = \frac14 \label{1}\eeq

\noindent where $\rho^2$ is the slope of the elastic ground state Isgur-Wise scaling
function, while the $\tau_{1/2}$, $\tau_{3/2}$ are the scaling
functions of the transition between the ground state and the $P$ wave ($L=1$) states with possible
radial excitation, $k = 1$ for the lowest $P$ state, $k = 2$, ~. for its radial
excitations. In the following, the $\tau_i$'s are always considered at the zero-recoil point
$w=1$. Therefore, to reduce notation, we {\it denote the corresponding values by} $\tau_{1/2}$, $\tau_{3/2}$. 

To demonstrate the sum rule, we shall not use the original Bjorken method starting from
commutators of field theory \cite{[bjorken]}, since our model does not contain a priori
such commutation relations. Rather, we shall calculate directly the sum of squares of
transition matrix elements to any final state and find that it has a very simple value,
independent of the dynamics. In the present approach, states have a fixed number of constituents $N$  (no pair creation). Moreover, for a confining interaction, they will be stable bound states~; this implies that saturation of the sum rule holds within the resonance-dominance and narrow-resonance approximations. \par

For the particular purpose of this article, it is found easier to treat on
equal footing the various states with different spins and which may be mesons or
baryons or states with any number of quarks ; we then abandon the manifestly covariant
formalism with Dirac spinors, and return to the initial bidimensional-spinor
formalism. In this formalism, current density matrix-elements write, without assuming $m_1$, $m'_1 \to \infty$~:

\[ <\vec{P}' |j^{\mu}| \vec{P}> = \int \prod_i d^3 p_i \sum_{s_is_1s'_1} \psi_{s'_1
s_i}^* (\vec{P}' - \Sigma \vec{p}_i, \{\vec{p}_i\} ) \]
\beq <\vec{p'}_1 s'_1|j^{\mu}|\vec{p}_1 s_1> \psi_{s_1 s_i} (\vec{P} - \Sigma \vec{p}_i , \{
\vec{p}_i\}). \label{2}\eeq
  
\noindent Index 1 denotes the active quark, while spectator quarks, from 2 to $N$, are denoted
by the generic index $i$. We disregard color and flavor. The $\psi$'s are what we call
the ``relative'' wave functions. They are obtained by dropping $\delta (\Sigma \vec{p}_i -
\vec{P})$ in the total wave function with total momentum $\vec{P}$ (we normalize any state
to $\delta (\vec{P}' - \vec{P})$ and 1 for any discrete index). \par

Part of our demonstration will be worked out without passing to the heavy quark limit. Let
us recall that at this stage the model is not covariant, but we do not need the
covariance property. For the first and most general demonstration of the sum rule, we do
not need either to make explicit the construction of the $\psi$'s from the internal wave
functions at rest. This construction shall be explained in detail in section 3, where one
really needs it. We need only to know that the $\psi$'s may be chosen to form a complete
orthonormal basis of the subspace with $\vec{P}$ fixed. Let us label it by a generic
label $n$ in addition to $\vec{P}$, we mean that~:

\beq\sum_{s_1s_i} \int \prod_i d^3p_i \ \psi_{n's_1s_i}^*(\vec{P} - \Sigma \vec{p}_i, \{
\vec{p}_i \}) \psi_{ns_1s_i} (\vec{P} - \Sigma \vec{p}_i, \{ \vec{p}_i\}) = \delta_{nn'} \
\ \ . \eeq

\noindent Then we have the closure relation~: 

\beq \sum_n \psi_{ns_1s_i}(\vec{P} - \sum_i \vec{p}_i, \{ \vec{p}_i \}) \psi_{n
{s'}_1 {s'}_i}^* (\vec{P} - \Sigma \vec{{p'}}_i,
\{\vec{{p'}}_i \}) = \delta_{s_1 {s'}_1} \prod_i \delta_{s_i {s'}_i}
\prod_i \delta (\vec{p}_i - \vec{{p'}}_i) \ \ \ .\label{4} \eeq

\noindent It is of course convenient to choose the basis among the eigenstates of
energy\break \noindent $\sqrt{M_{op}^2 + \vec{P}^2}$, and to choose for the label $n$ the labelling of the corresponding internal wave functions at rest $\varphi_n (\vec{k}_1, \{
{k}_i \})$, which will include, in addition to various internal spin and angular momentum
labels, excitation numbers. \par

Let us now consider the transition matrix-elements from one fixed state denoted as $n =
0$, $\vec{P}$, to all possible states, i.e. with all possible $n$, with another common
momentum $\vec{P}'$, induced by $j^{\mu}$. We choose for $j^{\mu}$ the elastic vector
current, although we could choose any other current and even any Dirac matrix ${\cal
O}$, with $m_1 \not= m'_1 \ $ - indeed, it is known that whatever the choice, we would obtain the
same Bjorken sum rule because of HQS. Our choice is made for transparency and
definiteness only. \par

With the closure relation eq.(\ref{4}) at hand, one can perform the following sum :

\beq h^{\mu \nu} (\vec{P}, \vec{P}') = \sum_n <n, \vec{P}'|j^{\mu}|0, \vec{P}> \times (<n,
\vec P'|j^{\nu}|0, \vec{P}>)^* \label{5} \eeq

\noindent which we will call the ``hadronic tensor'' in analogy with inclusive
leptoproduction or semi-leptonic decays. Let us stress that this sum should not be covariant
in principle, even if the theory were covariant, because each intermediate state has a
different energy at fixed $\vec{P}'$. We find trivially from eqs. (\ref{2} and \ref{4})~: 

\[ h^{\mu \nu} (\vec{P} , \vec{P}') = \sum_{s'_1 s_1 \widehat{s}_1 s_i} \int
\prod_{i=2}^N d^3 p_i \ \psi_{0\widehat{s}_1 s_i}^* (\vec{P'} - \sum_i \vec{p}_i, \{
\vec{p}_i \}) \psi_{0s_1s_i} (\vec{P} - \sum_i \vec{p}_i, \{ \vec{p}_i \})\] 
\beq  
(\bar{u}_{s'_1}(\vec{P}' - \Sigma \vec{p}_i) \gamma^{\mu} u_{s_1} (\vec{P} - \Sigma
\vec{p}_i)) 
  \times (\bar{u}_{s'_1} (\vec{P}' - \Sigma \vec{p}_i ) \gamma^{\nu}
u_{\widehat{s}_1}(\vec{P} - \Sigma \vec{p}_i))^* \ \ \ . \eeq

\noindent The important step already realized is to have reexpressed the hadronic
tensor in terms of the initial state ($n = 0$) wave function only. In fact, the
right-hand side of the equation is the average of a rather simple operator.
Introducing the Fourier transform of the current density~:

\beq \widetilde{j}(\vec{q}) \equiv \int d^3x \ e^{-i\vec{q}\cdot \vec{x}} \ j(\vec{x}) \eeq

\noindent one can write~:

\beq h^{\mu \nu} (\vec{P}, \vec{P}') = < 0 ; \vec{P} \left |
\widetilde{j}^{\nu}(\vec{q}) \ j^{\mu}(0)\right | 0; \vec{P}> \eeq

\noindent with $\vec{q} \equiv \vec{P}' - \vec{P}$.\par

Up to now all the written relations were valid in full generality. Let us now take
advantage of the heavy quark mass limit $m_1 \to \infty$. The important point is that in
this limit, the velocity of the active quark can be considered equal to the velocity of
the corresponding hadron, $\vec{v}$ or $\vec{v}'$. Then we can factor out of the
integral the spinor factors, which now depend only on $\vec{v}$ and $\vec{v}'$~:

\[ h^{\mu \nu}(\vec{v}, \vec{v}') = \sum_{s_1s'_1 \widehat{s}_1} \left [
\bar{u}_{s'_1} (\vec{v}') \gamma^{\mu} u_{s_1}(\vec{v}) \right ] \left [
\bar{u}_{s'_1}(\vec{v}') \gamma^{\nu} u_{\widehat{s}_1} (\vec{v}) \right ]^* \]
\beq \sum_{s_i} \int \prod_{i=2}^N d^3 p_i \ \psi^*_{0 \widehat{s}_1s_i} (\vec{P} -
\Sigma \vec{p}_i , \{ \vec{p}_i \}) \psi_{0s_1s_i}(\vec{P} - \Sigma \vec{p}_i, \{
\vec{p}_i \} ) \ \ \ . \eeq

\noindent A further simplification can be obtained if one averages as over the polarisation of the initial state 0 ; let us
call it $\lambda$. Then it is easy to convince oneself that~:

\beq \int \prod_{i=2}^N d^3 p_i {1 \over 2J_0 + 1} \sum_{s_i, \lambda}
\psi_{0 \widehat{s}_1 s_i}^{\lambda *} \ \psi_{0 s_1s_i}^{\lambda} \propto \delta_{s_1
\widehat{s}_1} \ \ \ . \eeq

First, this is easily seen to follow from rotation invariance if $\vec{P} = 0$. Then,
one can extend the result to arbitrary $\vec{P}$ using the following property of the
model~: in passing to wave functions in motion, the spectator spins $s_1$ are
simply rotated by Wigner rotations, which are cancelled by contraction, while for the
active quark spin $s_1$,$\widehat{s}_1$, the Wigner rotation reduces to unity in the $m_1 \to \infty$
limit. Then

\beq \int \prod_{i=2}^N d^3 p_i {1 \over 2 J_0+1} \sum_{s_i\lambda} \psi_{0  \widehat{s}_1
s_i}^{\lambda *} \ \psi_{0s_1s_i}^{\lambda} = \int \prod_{i=2}^N d^3p_i {1 \over 2} {1
\over 2J_0 + 1} \left ( \sum_{s_1 s_i \lambda} \psi_{0s_1s_i}^{\lambda *} \ \psi_{0s_1
s_i}^{\lambda} \right ) \delta_{s_1 \widehat{s}_1} \ \ \ . \eeq  

\noindent Whence our central relation~:
\[
 \bar{h}_{\mu \nu} (\vec{v} , \vec{v}') \equiv {1 \over 2J_0 + 1}
\sum_{\lambda} h_{\mu \nu}^{\lambda} (\vec{v}, \vec{v}') \]
\beq = {1 \over 2} \sum_{s_1, s'_1} \left [ \bar{u}_{s'_1}(\vec{v}') \gamma^{\mu} u_{s_1}
(\vec{v}) \right ] \left [ \bar{u}_{s'_1} (\vec{v}') \gamma^{\mu} u_{s_1}(\vec{v})
\right ]^* 
\equiv \bar{h}_{\mu \nu}^{free \ quark} \ \ \ . \label{12} 
\eeq

We have {\it exact duality} for the hadronic tensor, that is the general Bjorken
sum rule in the form of Isgur and Wise. Let us recall that it holds with any Dirac space
matrices instead of $\gamma^{\mu}$, $\gamma^{\nu}$. \par

Though it may appear trivial, it is worth stressing once more that such an exact duality
can hold only because {\it current matrix elements} are actually calculated as matrix elements
of a given operator between wave functions in motion that satisfy closure. If, on the
contrary, form factors are constructed according to phenomenological recipes, although
one may have at start orthonormal wave functions at rest, one cannot get duality for arbitrary $q^2$ or
$w = v \cdot v'$. Two properties of this heavy mass duality must also be underlined.
First, the limit of $h_{\mu \nu}$ is manifestly covariant, in contrast to $h_{\mu \nu}$
at finite $m_Q$. Second, this limit is the sum of the heavy mass limit of each product
of two matrix elements. This nonobvious exchange of limit and sum can be shown to hold
within the model, using closure for the wave functions in the heavy mass limit. That
entails that the sum of limits is convergent, which should not be necessarily true in general in field theory. \par

From this duality relation, one deduces easily the sum rule (\ref{1}) by reexpressing the current matrix elements in the definition of $h^{\mu \nu} (\vec{P}, \vec{P}') $ (eq.(\ref{5})) in terms of the standard invariant scaling functions, \cite{[isgur2]} . The deduction will be made by expanding $h_{00}$ to second order around $\vec{v}, \vec{v}' = 0$ ; for simplicity we choose $\vec v$ and $\vec v'$ collinear.
Moreover, we take as initial state a $0^-$ meson. One finds, taking into account that our own normalisation of states is
$\delta^3(\vec P-\vec P')$, different from the one of Isgur and Wise~:

\beq <0^- ; \vec{v}'|j^0|0^- ;\vec{v}> = {1 \over 2} {v_0 + v'_0 \over \sqrt{v_0 v'_0}}
\xi \approx \xi \approx 1 - {1 \over 2}	 \rho^2 (\vec v-\vec{v}')^2 \ \ \ . \label{13} \eeq

\noindent At this point, we make use of the fact, demonstrated in the preceding letter,
that the model indeed satisfies Isgur-Wise scaling, thanks in particular to the factorisation
of the wave function at rest into spin and rotationally invariant space
wave functions. This result extends to
radial excitations, by dropping the 1 in eq. (\ref{13}), because of orthogonality of wave functions at rest. On the other hand, we have not yet presented the
demonstration of covariance and scaling for the general case, which we postpone to
another paper. We assume for the moment this general result. Manifestly covariant and
scaling expressions will be presented in a forthcoming paper for the
transitions to $L = 1$ ($P$ wave states) \cite{[morenas]}. $0^-$ radial excitations yield ${\cal
O}((\vec{v}'-\vec{v})^2)$ amplitudes, therefore a negligible ${\cal
O}((\vec{v}'-\vec{v})^4)$ contribution to $h_{00}$. The other possible final states
through second order in $\vec v,\vec{v}'$ in the sum rule are $1^-$ and the $L = 1$ $\ 0^+$,
$1^+$, $2^+$ states, because $j^0$ is scalar under rotations. $0^- \to 0^+$ vanishes identically for
a vector current. $0^- \to 1^-$ and $2^+$ vanish since for $j^0$ they have the form
$\vec{v} \times \vec{v}'$ from covariance. Therefore, in addition
to the elastic $0^- \to 0^-$ contribution, we have just the $0^- \to 1^+$ one, to express in terms of
the scaling functions at zero recoil $\tau_i$. This is readily done~:

\beq <1/2, \vec{\varepsilon}_{\lambda}^{\ *}|j^0|0^-> \approx - \tau_{1/2} \
\vec{\varepsilon}_{\lambda}^{\ *} \cdot (\vec{v}' -\vec v)\quad (\lambda = 0) \label{14} \eeq

\beq <3/2, \vec{\varepsilon}_{\lambda}^{\ *}|j^0|0^-> \approx - \sqrt{2} \,\tau_{3/2}
\,\vec{\varepsilon}_{\lambda}^{\ *} \cdot (\vec{v}' -\vec v) \quad (\lambda = 0) \ \ \ .  \label{15} \eeq

\noindent We have made an expansion
only at the lowest, first order in $\vec v, \vec{v}'$, since these expressions appear in
squares (therefore we have made $v_0, v'_0 \approx 1$). Then~: 

\beq \sum_{\lambda} \left | <1/2, \lambda | j^0|0^-> \right |^2 + \sum_{\lambda} \left |
<3/2, \lambda | j^0|0^-> \right |^2 = \left ( |\tau_{1/2} |^2 + 2 |\tau_{3/2}|^2
\right ) (\vec v-\vec{v}')^2 \ \ \ . \label{16}\eeq  

\noindent Finally, summing on the P-wave states, including excitations~:

\beq h_{00}(\vec{v}, \vec{v}') \approx 1 - \rho^2 (\vec v-\vec{v}')^2 + \sum_k \left (
|\tau_{1/2}^k|^2 + 2 | \tau_{3/2}^k|^2 \right ) (\vec v-\vec{v}')^2\ \ \ . \eeq

\noindent On the other hand, the right-hand side of eq. (\ref{12}) is~:

\beq\bar{h}_{00}^{free \ quark} \approx 1 - {1 \over 4} (\vec v-\vec{v}')^2\eeq

\noindent whence the desired sum rule eq. (\ref{1}). \par

For baryons, the difference will be that the scaling functions for the elastic
transition are defined through the coefficient of $\gamma^{\mu}$ instead of $v^{\mu} +
v'^{\mu}$, whence~:

\beq < 1/2^+ |j^0|1/2^+> \sim 1 - {1 \over 2} \rho_B^2 (\vec v-\vec{v}')^2 - {1 \over 8} (\vec v-\vec{v}')^2 \eeq

\noindent because~:

\beq \bar{u}_{B'} \gamma^0 u_B \sim 1 - {1 \over 8} (\vec v-\vec{v}')^2 \eeq

\noindent with our normalisations ($u_B^+ u_B = 1$). In this case, the sum rule does not contain the $1/4$ term, but this has no deep physical meaning~:

\beq \rho_B^2 = \sum \ \hbox{inelastic contributions} \ \ \ . \eeq

One must note that one is dealing with the usual Bjorken sum rules, corresponding to the first term in the current commutators considered by Bjorken ("direct" contribution)~. It  has been shown by Bjorken that the direct and
$z$-graph contributions satisfy two separate, independent sum rules \cite{[bjorken]}. In our model as it stands, the $z$-graph contribution is absent due
to the absence of pair creation or annihilation, therefore the second sum rules are not present. But a slight modification of
the model, including pair creation or annihilation by the current and extending
correspondingly the space of states, would probably allow to get the commutation relations and to satisfy the second type of sum rules corresponding to the
$z$-graph contribution. \par \vskip 5 truemm

\section {Analysis in terms of internal wave
function matrix elements}

Now we pass to a more detailed calculation of the various contributions to the left-hand
side of the sum rule, as expressed in terms of internal wave functions at rest $\varphi$,
eigenstates of the mass operator. That is, we display the various effects due to the
hadron center-of-mass motion as treated relativistically. Then, we get more physical
insight. We use the expression (12) of the preceding paper \cite{[leyaouanc1]}, obtained by explicitation of
the relative wave function in terms of $\varphi$'s (this is before passing to the $m_Q \to
\infty$ limit)

\[ <\vec{P}\ ' |j^{\mu} |\vec{P}> = \int \prod_{i=2}^N d^3 p_i \times \sqrt{ {\Sigma
{p'}_{j}^{0} \Sigma p_{j}^{0} \over {M'}_{0} M_0}} \prod_{j=1}^N {\sqrt{{k'}_j^{0} k_j^0}
\over \sqrt{{p'}_{j}^{0} p_j^0}} \times \]
\beq   \sum_{s_j} \sum_{s'_j} \varphi '_{s'_j} (\{ \vec{k}'_i \})
\left [ {D} ({R'}_1^{-1})_{{s'}_1 {s'''}_1} \bar{u}_{{s'''}_1} \gamma^{\mu}
u_{{s''}_1} D(R_1)_{{s''}_1 s_1} \right ] 
 \prod_{i=2}^N D ({R'}_i^{-1}
R_i)_{{s'}_i s_i} \varphi_{s_j}(\{ \vec{k}_i \} ) \ \ \ .\label{22} \eeq

\noindent The $\vec{k}_i$ and $\vec{k}'_i$'s are the internal momenta corresponding
respectively to the initial and final hadrons. They are complicated functions of the
spectator momenta $\vec{p}_i$, obtained through the equations (4) of the preceding paper \cite{[leyaouanc1]}.
$p_{1,i}^0$ and $k_i^0$ denote simply $\sqrt{m_{1,i}^2 + \vec{p}_{1,i}^{\ 2}}$ and
$\sqrt{m_i^2 + \vec{k}_i^2}$ and $M_0 \equiv \sqrt{\Sigma k_i^2}$, $j$ runs from 1 to $N$.
$R_j$ are Wigner rotations defined in the same paper. Let us recall how these Wigner
rotations are practically related to boosts. Let $\Lambda$ be a Lorentz boost~:

\beq \sqrt{p^0} \ \Lambda \ u_s (p) = \sqrt{(\Lambda p)^0} \ D_{ss'}(R) \ u_{s'} (\Lambda
p) \eeq

\noindent with the rotation $R$ defined by~:

\beq R = B^{-1}_{\Lambda p} \Lambda B_p \eeq

\noindent $B_a$ is the boost which applies $(\sqrt{a^2},\vec 0)$ on the four vector $a_\mu$. The root factors in
the formula are needed because of our normalization of states. From this formula, one
can deduce easily $D_{ss'}(R)$ knowing $\Lambda$ and $p$. \par

Let us identify the origin of the various factors involved in eq. (\ref{22}), in addition to
the wave functions $\varphi$'s and to the free quark current density. In front of $\varphi '$, there is the product of 
square roots of the Jacobians corresponding to the change of variables $\vec{p}_1, \{
\vec{p}_i \} \to \{ \vec{k}_i \}$, $\vec{P}$ and $\vec{p}\ '_1, \{ \vec{p}_i \} \to \{
\vec{k}_i \}, \vec{P}'$ ; the presence of these factors ensures unitarity of the
cor\-res\-pon\-ding functional transformations. Between $\varphi '$ and $\varphi$, the Wigner rotations of quark spins ensure the passage from the ordinary one-particle
spins to the internal spins ; note that the $D$ matrices are unitary by themselves. \par

In the limit $m_Q \equiv m_1$ or $m'_1 \to \infty$, the expression simplifies
considerably. For convenience, we choose $\vec{v}$ and $\vec{v}'$ along the same axis $0z$ and
count them by the algebraic numbers $v_z$ and $v'_z$. We maintain
$\vec{v} \not= 0$ to check the requirements of covariance on matrix elements, as imposed by the expressions (\ref{13}), (\ref{14}), (\ref{15}). First, the relation between the $k_i$'s and $p_i$'s becomes~:
\beq
k_i^0 = v^0 p_i^0 - v_zp_{iz}, \qquad
k_{iz} = v^0 p_{iz} - v_zp_i^0, \qquad
\vec{k}_{iT} = \vec{p}_{iT}, 
\eeq 

\noindent where $T$ denote the component perpendicular to $0z$. One has an analogous
relation for the ${\vec{k}'}_i$'s with $v'$ instead of $v$. Then, the Jacobian factors take the very simple
form~:

\beq \prod_i {\sqrt{v^0p_i^0 - v_z p_{iz}} \sqrt{v'^0 p_i^0 - v_z' p_{iz}} \over p_i^0} \ \ \
. \eeq

\noindent The Wigner rotations of the active quark tend to unity because its momentum
becomes parallel to the direction of the boost, which becomes $\vec{v}$ or $\vec{v}'$.
Finally, the active quark current density tends to~:

\beq \bar{u}_{s'_1} \ \gamma^0 \ u_{s_1} \approx \left [ 1 - {1 \over 8} (\vec v - \vec v')^2 \right ]
\delta_{s'_1 s_1} \eeq

\noi and can be once more factored out of the integral. Whence the final expression in
the limit $m_Q \to \infty$

\[ <\vec{P}'|j^{\mu}|\vec{P}> = \sum_{s'_1s_1} \bar{u}_{s'_1} \gamma^{\mu} u_{s_1} \int
\prod_{i=2}^N d^3 p_i \prod_{i=2}^N {\sqrt{(p_i\cdot v) (p_i \cdot v')} \over
p_i^0} \]
\beq \sum_{s_is'_i} \varphi_{s'_1 s'_i} (\{\vec{k}'_i \}) \prod_{i=2}^N D ({R'}_i^{-1}
R_i)_{s'_is_i} \ \varphi_{s_1s_i}(\{ \vec{k}_i \}) \ \ \ . \eeq

\noi Let us now expand the various factors in this expression around $\vec v$, $\vec v' = 0$, at
fixed $\vec{p}_i$, $\vec{k}_i$ being a function of $\vec{p}_i$ and $\vec v$. We note that at $\vec v
=\vec  v' = 0$, one gets simply the scalar product of the internal wave functions at rest
$\varphi$, $\varphi '$. Therefore it is 1 for the elastic transition $0
\to 0$ and 0 for inelastic transitions. 

Then, to calculate the $\tau_i$ 's at zero recoil, we have once more to calculate
inelastic amplitudes only through lowest, i.e. first order ${\cal O}(\vec v)$, or ${\cal O}(\vec v')$. Let us first calculate these inelastic contributions. We have the following
lowest order expansion of the integral~:

\[ \left . <n|j^0|0> = \int \prod_i d^3 p_i \sum_i \varphi '^*(\vec{p}_i) \left [
(v'_z - v_z) \left (  p_i^0 {\partial \over \partial p_{iz}} + {\partial \over \partial
p_{iz}} p_i^0 \right ) \right / 2 \right . \]
\beq \left . + {i \over 2} (v_z' - v_z) { (\vec{\sigma} \times
\vec{p}_{iT})_z \over p_i^0 + m_i} \right ] \varphi (\vec{p}_i) \eeq

\noindent where scalar product on spin space is implied. The first term comes from the combination of the variation of the Jacobian factors,
and the variation of the argument $k$ of the wave function, the second one from Wigner
rotations. Each has separately a factor $v_z' - v_z$, as required by
covariance. Indeed, the covariance requires a factor $(v'_z - v_z)$ for the sum ; now, if we had dropped the spin and considered scalar quarks, the first
term only would remain ; covariance then requires a factor $v_z' - v_z$ on this factor
separately. Note that in this approximation the active quark density
$\bar{u}_{s_1} \gamma^0 u_{s_1} \approx 1$ does not give a contribution. We can
replace $i \ \partial/\partial p_{iz}$ by the more suggestive notation $z_i$ ; it
is indeed the space coordinate operator. In addition, we
henceforth denote matrix elements between $\varphi_n$, $\varphi_{n'}$ as
($n'||n$). Then~:

\beq <n|j^0|0> \approx (v_z' - v_z) \sum_i \left ( n \left | -{ p_i^0 iz_i + iz_i p_i^0
\over 2} + {i \over 2} {(\vec{\sigma}_i \times \vec{p}_{iT} )_z \over p_i^0 + m} \right
| 0 \right ) \ \ \ . \eeq

\noindent If we particularize to mesons, $i = 2$ only ; we can drop
the index $i$ and write~:

\beq <P_{wave}|j^0|S_{wave}> \approx (v_z' - v_z) \left ( n\left | -{p^0 iz + iz p^0 \over
2} + {i \over 2} {(\vec{\sigma} \times \vec{p}_T)_z \over p^0 + m} \right |0 \right ) \ \
\ . \eeq 

\noindent We now particularize to a $0^-$ initial state, and  to $n = 1^+$. Then, one finds from
the identification with eq. (\ref{16}):
\beq \sum_k \left ( \left | \tau_{1/2}^k \right |^2 + 2 \left | \tau_{3/2}^k \right |^2
\right ) =  \sum_{1/2, 3/2} \sum_k \left | \left ( 1^+, k \left | - {p^0iz + iz p^0 \over 2} + {i
\over 2} {\left ( \vec{\sigma} \times \vec{p}_T \right )_z \over p^0 + m} \right |
0^- \right ) \right |^2 \ \ \ . \eeq

\noindent It is clear that the operator on the right-hand side leads only to
transitions to $1^+$ (it has $L = 1)$. Therefore we can as well replace the
summation on states by a summation on the whole Hilbert space, and using closure of the
$\varphi$'s, we end with~: 

\beq \sum_k \left ( \left | \tau_{1/2}^k \right |^2 + 2 \left | \tau_{3/2}^k \right |
\right )^2 =  \left ( 0^- \left | \left ( {p^{0}z + zp^0 \over 2} \right )^2 + {1 \over
4} \left [ {\left ( \vec{\sigma} \times \vec{p}_T \right )_z \over p^0 + m} \right ]^2
\right | 0^- \right ) \ \ \ . \eeq

\noindent We have taken into account that $\vec{\sigma}$ average on $S = 0$ states
is 0, so that the two contributions add in squares. Further simplification is obtained by using rotation invariance in the Wigner rotation
contribution~:

\beq \sum_k \left ( \left | \tau_{1/2}^k \right |^2 + 2 \left | \tau_{3/2}^k \right |^2
\right ) = \left ( 0^- \left | \left ( {p^0z + zp^0 \over 2} \right )^2 + {1 \over 6}
{\vec{p\,}^{2} \over (p^0 + m)^2} \right | 0^- \right ) \label{34}\eeq

\noindent which is an average on space wave functions only. \par

As to $\rho^2$,
we could simply refer to the preceding paper, eq. (29) of the published version\cite {[leyaouanc1]}. However, it is more instructive physically to
recalculate $\rho^2$ along the same lines as we have just done for the $\tau$'s, i.e. directly
in the formalism with bidimensional spin and Wigner rotations. We have just to push the
expansion in $\vec{v}$ and $\vec{v}'$ of the matrix element through second order. It
happens that the result can be decomposed in a manner
similar to the $\tau$'s. One has three types of effects which make the matrix element depart from its zero recoil value, 1~: \par
i) one from the Jacobians and the spatial wave function arguments variations \par
ii) one from the Wigner rotations \par
iii) one from the current the active quark, which was not present for P waves. \par

\ noi Covariance requires the whole result to be of the form $1 - (\rho^2/2) (\vec v' - \vec v)^2$. But in fact each effect gives a separate
contribution, which is  $\propto (\vec v' - \vec v)^2$. Indeed for iii) it is obvious, see eq. (\ref{39}) below (it is the requirement of covariance for a one quark
state). Then the effects i) and ii) must also combine to give a contribution of this form. Now, there are no crossed terms between them ; these should correspond to the
product of factors of first order in the velocity~: first
order terms from the Wigner rotation effect ii) contain a spin operator $\vec{\sigma}$, which averages to 0 on a $0^-$ state~; then, these first order terms cannot
combine with one from i) (the latter effect does not generate spin operators). i) and ii) give therefore non-interfering additive contributions. Moreover each one must have the form $(\vec v' - \vec v)^2$. For the first effect i), it is seen by dropping spin and
considering scalar quarks~: then i) would give the only contribution and therefore it must be
$\propto (\vec v' - \vec v)^2$ separately by covariance. It results that the same form must holds for ii). The contribution i) to $\rho^2$ is found to be, by a
somewhat lengthy calculation~:

\beq \rho^2_{space} = \left ( 0 \left | \left ( {p^0z + z p^0 \over 2} \right )^2 \right |
0 \right ) \ \ \ . \eeq

\noi The contribution ii) is

\beq \rho^2_{wigner} = \left ( 0 \left | {1 \over 6} {\vec{p}^{\ 2} \over (p^0 + m)^2}
\right | 0 \right ) \ \ \ . \eeq

\noindent Finally, from the active quark current~:

\beq<p'_1 s'_1 |j^0|p_1 s_1> = 1 - {1 \over 8} (\vec v' - \vec v)^2 \label{39} \eeq

\noi one has~:

\beq \rho^2_{dirac} = {1 \over 4} \eeq

\noindent whence~:

\beq \rho^2 = \rho^2_{space}+\rho^2_{wigner}+\rho^2_{dirac}=
\left ( 0 \left |
\left ( {p_0z + zp_0 \over 2} \right )^2 + {1 \over 6} {\vec{p}^{\ 2} \over (p_0 + m)^2}
\right | 0 \right ) + \frac14 \eeq

\noi BIW sum rule is obviously satisfied, considering eq. (\ref{34}) .

It is seen that ``spatial'' contributions to $\rho^2$ and $\tau^2$'s, i.e. the one
obtained by dropping quark spin, cancel each other as it should since, for {\it scalar}
quarks, the sum rule should write~:

\beq \rho^2 - \sum_k \left | \tau^k \right |^2  = 0 \eeq

\noi since in this fictitious case $h_{00}^{free \ quark} \approx 1$ instead of $1 - 1/4 (\vec v - \vec v')^2$. Wigner rotation contributions cancel also each other, which is seen to be due to the
unitarity of Wigner rotations. \par 
Only the 1/4 coming for the active quark
current has no counterpart in the $\tau^2$'s. One notes once more that this 1/4 is present in the
$j^0$ matrix element of any elastic transition. It {\it will be present also
for a free quark}. Let us also recall that the
absence of 1/4 in baryon sum rules has nothing physical~: it is due to the definition
of invariant form factors, which are differently related to $j^0$ matrix elements. In
terms of the latter, the only difference between mesons and
baryons comes from the fact that $i$ sums run over two spectator quarks instead of
one~; but this refers to what we can term the {\it compositeness contribution} $\rho^2_{space} + \rho^2_{wigner}$.\par
In the following two subsections, because of the BIW sum rule, one needs only discuss $\rho^2$, from which the parallel comments
on the inelastic contributions can be deduced trivially.

\subsection{Non relativistic expansion}

It is useful to comment briefly on the order of magnitude of the contributions in a
nonrelativistic expansion, i.e. in powers of $v/c$ where $v/c$ is now the
{\it internal} velocity of the {\it light} quarks. We know that this velocity is not actually small~; nevertheless, this expansion  gives more
physical insight. The dominant contribution to $\rho^2$ is
the first one $\rho^2_{space}$, from the spatial wave function. One finds at lowest
order~:

\beq \rho^2 \approx \rho^2_{space} \approx m^2(0|z^2|0) = {\cal O}\left (
(v^2/c^2)^{-1} \right ) \eeq

\noindent $= {m^2R^2 \over 2}$ in the h.o. case. This corresponds for $\tau$ to the
use of the usual dipole formula. The Wigner rotation contribution is on the contrary
highly suppressed~:

\beq \rho^2_{wigner} \approx {1 \over 24} \left ( 0 \left | {\vec{p}^{\ 2} \over m^2}
\right | 0 \right ) = {\cal O} (v^2/c^2) \ \ \ . \eeq

\noindent $\rho^2_{dirac} = 1/4$ is in between~: ${\cal O} \left (
(v^2/c^2)^0 \right )$~; then it is actually the dominant relativistic effect coming from
spin. This contrasts with Ref. \cite{[close]} ; there, the discussion identifies the spin effect with the Wigner rotation effect only.
Denoting ground state averages as $( \ )_0$, we can also write a $v/c$ expansion of the
full $\rho^2$~:

\beq \rho^2 = m^2 (z^2)_0 + {1 \over 2} \left ( \vec{p}^{\ 2}z^2 + z^2 \vec{p}^{\ 2} \right
)_0 + \frac34 + {\cal O}(v^4/c^4) \label{47}\eeq

\noindent where the second term is ${\cal O}\left ( (v^2/c^2)^0 \right )$ and comes
for the ``spatial'' contribution $\rho^2_{space}$. It is seen that the lowest order
relativistic corrections cannot be reduced in general, to the famous 1/4, as seems
often assumed. Moreover they do not come from spin only~;

\beq \rho^2_{space} = m^2 \left ( z^2\right )_0 + 1/2 \left ( \vec{p}^{\ 2}z^2 +
z^2 \vec{p}^{\ 2} \right )_0 + \frac14 + {\cal O}(v^4/c^4) \eeq

\noindent contributes to them as observed in Ref. \cite{[close]}. Finally these
additional relativistic corrections are not independent of the potential. Indeed
$(\vec{p}^{\ 2} z^2 + z^2 \vec{p}^{\ 2})_0$ has not even a definite sign. \par

Truly, it is a convention to take $m^2(z^2)_0$ as the starting
point of the $v/c$ expansion. Indeed, it must not be forgotten that the average are
taken on internal wave functions, which in realistic cases shall include by themselves
{\it relativistic binding corrections}. Consequently, since eq.(\ref{47}) does not
make explicit the latter relativistic effects, it is not the
full $v/c$ expansion around a truly nonrelativistic limiting case. Nevertheless, the relativistic corrections written
in eq. (\ref{47}) may be considered as a reasonable answer to the question inasmuch as we are
concerned only with relativistic corrections to $\rho^2$ due to the center-of-mass motion
of hadrons. 

\subsection{The lower bounds}

It is obvious that in this model there is no upper bound since
$\rho^2$ can be arbitrarily large in the nonrelativistic regime, e.g. ${m^2R^2
\over 2}$ for an harmonic oscillator. On the other hand, we have found in the
preceding paper\cite{[leyaouanc1]} a lower bound $\mathrm {Inf} \ \rho^2 = 3/4$, which is larger than 1/4~, the famous Bjorken lower
bound. To investigate further the compositeness contribution $\rho^2_{space} + \rho^2_{wigner}$, it reveals useful to
discuss the meaning of this bound. One may wonder why one cannot reach 1/4. Indeed, it
would be tempting to suppose, by reverting the above argument, that one could
reduce arbitrarily the ``compositeness'' contribution by reducing $m^2(z^2)_0$, which corresponds to going to a highly
relativistic situation ($R^2 \to 0$). But the idea is wrong, due to the relativistic
effects. Part of the reason is the Wigner rotation contribution $\rho^2_{wigner}$, which,
for $|\vec{p}| \gg m$, amounts to 1/6. But another reason lies in the corrections to the ``spatial''
contribution $\rho^2_{space}$. \par

Indeed, when $m^2 (z^2)_0$ is reduced by going to large average momenta, $p^0$ is
increasing in average and then, the final outcome is that the full expression of $\rho^2_{space}$, $\left ( 0 \left | \left
( {p^0z + zp^0 \over 2} \right )^2 \right | 0 \right )$, is bounded from below. We
can see this by the change of variable introduced in \cite{[leyaouanc1]}, such as $dx = p^0 {d \over dp}$ ~:

\beq p/m = \mathrm{sh} \ x \ , \ p^0/m = \mathrm{ch} \ x \eeq

\beq x = \mathrm{Argsh} \ p/m = \mathrm{Argch} \ p^0/m \eeq

\noindent $x$ coincides nonrelativistically with $p$, but behaves as $\log p/m$
for large $p = |\vec{p}|$. We can write $\rho^2_{space}$ as~:

\beq \left | \left | {p^0z + zp^0 \over 2} \varphi_0 \right | \right |^2 \eeq 

\noindent where $\varphi_0$ is the space part of the $0^-$ wave function. Then,
using rotation invariance of $\varphi_0$

\beq 
\left | \left | {p^0z + zp^0 \over 2} \varphi_0 \right | \right |^2 = {1 \over
3} \left | \left |  \left ( p_0 {d \over dp} + {1 \over 2} {p \over p_0} \right ) \varphi_0
\right | \right |^2 = {1 \over 3} \int d^3 p \left | \left ( {d \over dx} + {1 \over 2} \mathrm{th} \ x
\right ) \varphi_0 \right |^2 \ \ \ .
\eeq

\noindent Since $d^3 p = 4 \pi \ \mathrm{sh}^2x \ \mathrm{ch} \ x \ dx$ on spherically symmetric
wave functions, we define~:

\beq g = \sqrt{4 \pi} \ \mathrm{sh} \ x \sqrt{\mathrm{ch} \ x} \varphi_0 (m \ \mathrm{sh} \ x) \eeq

\noindent one then finds~:

\beq  \left | \left | \left ( p_0 {d \over dp} + {1 \over 2} {p \over p_0} \right )
\varphi_0 \right | \right |^2_{\vec{p}} =  \left |
\left |   \left ( {d \over dx} - \mathrm{th} \ x \right ) g \right | \right |_x^2 
 =  \int dx \ g \left ( 1 - {d^2 \over dx^2} \right ) g = 1 + \left | \left | {dg \over dx} \right | \right |_x^2 \ \ \ . \eeq

\noindent The second term is positive and has 0 as lower bound, (obtained by spreading
in\-de\-fi\-ni\-tely $g(x)$). Therefore

\beq \mathrm {Inf} \left ( 0 \left | \left ( {p_0z + zp_0 \over 2} \right )^2 \right | 0 \right
) = \frac13 \ \ \ . \label{55} \eeq

\noindent Note that $\varphi_0$ can be chosen arbitrary because the mass operator is
arbitrary, except that $M_0 > 0$. Since the Wigner rotation contribution is positive
and smaller than 1/6 $\ $(since $p/p_0 + m \leq 1$), one gets~: $7/12 \leq {\mathrm {Inf}}
\rho^2 \leq 9/12 = 3/4$. The more complete argument of \cite{[leyaouanc1]} is necessary
to show that~: \beq\mathrm {Inf} \ \rho^2 = \frac34  \eeq 

\noindent exactly, i.e. the Wigner rotation attains its maximum at the lower bound of
$\rho^2$. This reflects the fact that the bound is attained at large average of
$|\vec{p}|$. \par

Since $\rho^2_{space}$ is seen to be a major contribution to the lower bound
3/4 = 1/3 + 1/6 + 1/4, it is worth specifying its origin. First it is
an effect of compositeness, along with Wigner rotations. More specifically, it reflects
the pure effect of {\it spatial} extension of the bound state. Finally, still
more specifically, the bound 1/3 in eq. (\ref{55}) reflects the effect of the { \it relativistic}
transformation on the spatial wave functions. Indeed, nonrelativistically $\rho^2_{space}$ ($m^2(z^2)_0$) could be made arbitrarily small . The effect of the relativistic
transformation law amounts practically to replacing $m$ by the larger $p^0$ (as noticed in \cite{[close]}). Then the contribution is bounded from below as just
demonstrated, and in accord with the qualitative guess $p^0 \sim |\vec{p}|$ for large
$|\vec{p}|$ and $|\vec{p}| |z| \sim 1$. \par \vskip 5 truemm

\section {Conclusion}

In conclusion, we see that the heavy quark limit of quark models for currents based on
the Bakamjian-Thomas construction of states exactly satisfies the important sum rule or
duality relation discovered by Bjorken and further analyzed by Isgur and Wise. This
property is essentially due to the nontrivial fact that the wave functions in
{\it motion} satisfy a closure relation. This, added to the previous demonstration
that it is covariant and scales as required by the Isgur-Wise heavy quark symmetry
relations, increases the interest of the model. Another aspect illustrated by the present
analysis is the capacity of quark models to give a physical insight in the saturation of
general field theoretic relations through the use of the concepts of bound state physics.
Finally, one is stimulated to investigate the subdominant  regime in $1/m_Q$ which is
involved in other important sum rules such as the Voloshin ``optical'' sum rule ; in
general, properties such as covariance or the conservation of the current are lost, but
could perhaps be recovered under certain conditions. \vskip 5 truemm

\noindent {\bf Acknowledgment.} We would like to thank J. F. Mathiot for having drawn
our attention to recent work connected with the Bakamjian-Thomas formalism. This work was supported in part by the CEC Science Project
SC1-CT91-0729 and by the Human Capital
and Mobility Programme, contract CHRX-CT93-0132.\par

\end{document}